\documentclass[prl,aps,twocolumn]{revtex4}
\begin{document}

\title{\bf Vertex Operators of Open String States in the Intersecting D-brane World}
\author{ Huan-Xiong Yang 
         Mingxing Luo    
         and Wei-Shui Xu}
\affiliation{Zhejiang Institute of Modern Physics, Department of Physics, Zhejiang University, Hangzhou,
Zhejiang 310027, P. R. China \\
E-mail: hxyang, luo, wsxu@zimp.zju.edu.cn}
\date{\today}

\begin{abstract}
Starting with a new bosonization scheme for the $\beta\gamma$ CFT of the super-conformal ghosts, 
vertex operators are constructed for massless open string states in the intersecting D-brane world.
These vertex operators satisfy all requirements for a consistent RNS formulation of superstring theories,
so GSO projections can be properly taken.
\end{abstract}

\maketitle

With the advent of D-branes, 
studies have been intensified for phenomenologies of Type I/II string theories in recent years. 
D-brane configurations permit chiral fermions 
and the gauge group structure $U(3)\times U(2) \times U(1)$. 
Standard-like models and their supersymmetric extensions have been constructed from
orbifolds or orentifolds of Type II string theories with D-branes\cite{1,2}, 
in particular from the intersecting D-brane world\cite{3}. 
In these models, GSO projections, which are closely related to vertex operators,
are crucial to maintain the chiralities of massless states in the Ramond (R) sector 
as well as the $N=1$ supersymmetry.
However, a proper definition of vertex operators seems to be wanting 
for open strings connecting D-branes which are intersecting with nontrivial angles\cite{Douglas}.

Vertex operator is a concept of 2-dimensional conformal field theories (CFTs). 
They play an important role in superstring theories and provide elegant
(conformally invariant) means to calculate string interactions. 
Fermion vertex operators, in particular, are needed to define the worldsheet spinor number
and GSO projections\cite{FMS}.
They help to expose the space-time supersymmetries 
in the manifestly Lorentz invariant Ramond-Neveu-Schwarz (RNS)
formulation of superstring theories. 
By construction, vertex operators turn physical states into physical states. 
Fermionic vertex operators $V_{F}$ convert fermions into bosons and vice versa, 
which form spinors of the Lorentz group $SO(1,\,p)$ $(p\leq9)$. 
Most importantly, vertex operators should have a conformal dimension of {\it one}. 
Otherwise,
vertex operators cannot turn a physical state into another physical state 
and a covariant quantization cannot be formulated for superstring theories. 
The realization of these conditions could be challenging, 
but can be greatly simplified by the ``bosonization'' 
technique\cite{Coleman,Mandelstam}.

In Type I superstring theories with D9-branes, where open strings obey only
Neumann-Neumann boundary conditions, the construction of vertex operators
was based upon the so-called FMS bosonization scheme 
of the $\beta\gamma$ CFT of the super-conformal ghosts\cite{FMS}. 
The vertex operator of the super-conformal ghost ground state,
with conformal dimension $3/8$ in the R sector and $1/2$ in the NS sector, 
provided the long-missing pieces to the full vertex operators with the
correct conformal dimension.
Introducing nontrivial D-branes into Type I/II superstring theories, 
open strings can also obey Neumann-Dirichlet or Dirichlet-Neumann conditions. 
A naive generalization of the FMS prescription,
however, does not yield qualified vertex operators.
The failure can be illustrated by an
example of the supersymmetric $``n=2"$ case with D-branes intersecting at angles\cite{Douglas}. 
The brane configuration consists of two 2-branes in four dimensions $Z_{3}=X_{6}+iX_{7}$
and $Z_{4}=X_{8}+iX_{9}$. 
The first 2-brane is oriented along the $X_{6,\,8}$-axes.
The second one is obtained by rotating the first one by an angle $\alpha\pi$ in the
$Z_{3}$ plane and $-\alpha\pi$ in the $Z_{4}$ plane
($0 \leq \alpha \leq \frac{1}{2}$). 
For massless open strings connecting two such 2-branes, 
a straightforward generalization of the FMS scheme would result in the following vertex operators
\begin{eqnarray}\label{eq: 1}
e^{i(1-\alpha)H_3}e^{i\alpha H_4}e^{-\phi}\,, \nonumber \\
e^{-i\alpha H_3}e^{-i(1-\alpha) H_4}e^{-\phi}\,,
\end{eqnarray}
in the NS sector\cite{Douglas}. Here the worldsheet fermions $\Psi^{\mu}(z)$ are
bosonized by free holomorphic scalars $H_{a}(z)$\cite{Pol} 
and $e^{-\phi}$ is from the super-conformal ghosts.
The vertex operators in Eq.(\ref{eq: 1}) have conformal dimension $h=1-\alpha + \alpha^2$,
which is not equal to {\it one} unless $\alpha$ vanishes. 
Constructions in the R sector yield similar results.
A close inspection shows that the mismatch can only be due to the super-conformal ghost part. 
The factor $e^{-\phi}$ does not
reflect the influence of the open string boundary conditions correctly.

In this letter, we provide a new bosonization scheme of the $\beta\gamma$ CFT.
Based upon this, we construct a set of new ghost vertex operators, which provide the
needed extra pieces to the full vertex operators of open string states 
with supersymmetries\cite{Douglas}. 
These full vertex operators satisfy all 
requirements for a consistent RNS formulation of superstring theories,
so that GSO projections can be properly taken.

{\it An Alternative Bosonization of the $\beta\gamma$ CFT.} 
We are interested in the $\beta\gamma$ CFT from the BRST quantization of superstring theories, 
where $\beta(z)$ and $\gamma(z)$ are Faddeev-Popov super-conformal ghosts from gauge-fixing the superstrings. 
The classical action of the $\beta\gamma$ CFT is,
\begin{equation}\label{eq: y1}
S=\frac{1}{2\pi}\int d^2 z \beta \bar{\partial}\gamma \,\,.
\end{equation}
$\beta$ and $\gamma$ are commuting fields and could be regarded as ``bosons'' in
2-dimensional Euclidean space. They have the following operator product expansions
(OPEs),
\begin{equation}\label{eq: y2}
\left.
\begin{array}{ll}
 \beta(z)\gamma(0)  = - \frac{1}{z} + O(1)\,,
\,\,\,\,& \gamma(z)\beta(0) = \frac{1}{z} +O(1)\,, \\
 \beta(z)\beta(0)   = O(1)\,,
& \gamma(z)\gamma(0) = O(1) \,.
\end{array}\right.
\end{equation}
In general, a conformal invariant system may have more than one conserved
energy-momentum tensor. To give the conformal dimensions,
 $h_\beta=3/2$ and $h_\gamma= -1/2$, respectively, 
as required by the BRST quantization procedure,
the energy-momentum tensor of the $\beta\gamma$ CFT is uniquely defined as,
\begin{equation}\label{eq: y3}
T_{\beta\gamma}(z)= :[\partial \beta(z)]\gamma(z): -
\frac{3}{2}:\partial[\beta(z)\gamma(z)]:,
\end{equation}
so the central charge of the theory is $c_{\beta\gamma}=11$.

Though $\beta(z)$ and $\gamma(z)$ are bosonic, they need to be further ``bosonized''
in order to conveniently construct the vertex operators of ghost ground
states\cite{Pol}. By bosonization it is meant that these fields are re-expressed as
operator exponentials of other boson fields. 
In the FMS scheme\cite{FMS}, 
the bosonization was realized in terms of a holomorphic scalar $\phi(z)$
and a pair of anti-commutating bosonic ghosts $\xi(z)$ and $\eta(z)$,
\begin{eqnarray}\label{eq: y4}
\beta(z) \cong \exp[-\phi(z)]\partial\xi(z), \,\,\,\,& \gamma(z) \cong
\exp[\phi(z)]\eta(z).
\end{eqnarray}
These ``bosons'' form two separate 2-dimensional CFTs. 
In the $\phi$ CFT, the
fundamental OPE is $\phi(z)\phi(0)\sim -\ln z$ and the energy-momentum tensor is,
\begin{equation}\label{eq: y5}
T_{\phi}(z)= -\frac{1}{2}:[\partial \phi(z)][\partial \phi(z)]: - \partial^2 \phi(z).
\end{equation}
$\xi(z)$ and $\eta(z)$ are decoupled from $\phi$ and form an anti-commutating CFT\cite{Pol}. 
The fundamental OPEs of the $\xi\eta$ CFT are postulated to be
\begin{equation}\label{eq: y6}
\left.
\begin{array}{ll}
\xi(z)\eta(0) = \frac{1}{z} + O(1)\,, \,\,\,& \eta(z)\xi(0) = \frac{1}{z} +O(1)\,,  \\
\partial\xi(z)\partial\xi(0)= O(z)\,, & \eta(z)\eta(0) = O(z),
\end{array}
\right.
\end{equation}
and the energy-momentum tensor is 
\begin{equation}\label{eq: y7}
T_{\xi\eta}(z)= - \eta(z)\partial\xi(z).
\end{equation}
As a result, 
the ``bosonic ghosts'' $\eta(z)$ and $\xi(z)$ have conformal dimensions $h_{\eta}=1$
and $h_{\xi}=0$, respectively.
Of course, $T_{\beta\gamma}(z) \cong T_{\phi}(z) + T_{\xi\eta}(z)$.

To construct fermion vertex operators of the open states in the R sectors, 
the FMS bosonization of the $\beta\gamma$ CFT is only useful
when the full vertex operators form a spinor of the Lorentz group $SO(1,9)$. 
In Type II superstrings, open string states arise along with the appearance of BPS D-branes.
In these theories, open strings obey Neumann boundary conditions in directions
parallel to the D-brane's worldvolume while Dirichlet conditions in directions
perpendicular to the worldvolume. 
The open string fermion vertex operators are
generally spinors of the Lorentz group $SO(1,p)$ with $p\leq 9$. 
For $p<9$, the FMS scheme fails. 
As mentioned above, the conformal dimension of vertex operators
based upon the FMS bosonization does not equal to {\it one} in general. 
Technically, the failure is due to the fact 
that the number of independent free bosons in FMS bosonization is too small 
and there are no adjustable parameters to satisfy the $h=1$ requirement.

We now propose an alternative bosonization scheme for the $\beta\gamma$ CFT. Instead
of one free holomorphic boson, we introduce four and
postulate the following equivalence relations,
\begin{eqnarray}\label{eq: y8}
& \beta(z) & \cong \exp[-\phi_{1}(z) -\phi_{2}(z) - \frac{1}{\sqrt
2}\chi_{1}(z)-\frac{1}{\sqrt 2}\chi_{2}(z)]\partial\xi(z) , \nonumber \\
& \gamma(z)& \cong \exp[ \,\,\,\,\phi_{1}(z) +\phi_{2}(z) + \frac{1}{\sqrt
2}\chi_{1}(z)+\frac{1}{\sqrt 2}\chi_{2}(z)]\eta(z) ,
\end{eqnarray}
$\xi(z)$ and $\eta(z)$ are same as those in the FMS bosonization, which obey the OPEs
in Eq. (\ref{eq: y6}) and have the energy-momentum tensor in Eq. (\ref{eq: y7}).
$\phi_{i}(z)$ and $\chi_{i}(z)$ ($i=1,\,2$), which are
decoupled from the $\xi\eta$ CFT and independent of each other, have the following OPEs,
\begin{equation}\label{eq: y9}
\left.
\begin{array}{lll}
\phi_{i}(z)\phi_{j}(0) \sim - & \delta_{ij} \ln z \,,\,\,\,& \\
\chi_{i}(z)\chi_{j}(0) \sim   & \delta_{ij} \ln z \,,& (i,\,j=1,\,2)\,
\end{array}
\right.
\end{equation}
and energy-momentum tensors,
\begin{equation}\label{eq: y10}
\left. \begin{array}{lll} T_{\phi_i}(z) = -& \frac{1}{2}:[\partial \phi_i(z)][\partial
\phi_i(z)]: - \partial^2 \phi_i(z), \,\,& \\
T_{\chi_i}(z) = & \frac{1}{2}:[\partial \chi_i(z)][\partial \chi_i(z)]: + \omega_{i}
\partial^2 \chi_i(z)\,, &(i = 1, \,2) \\
\end{array}
\right.
\end{equation}
with constants $\omega_{1}$ and $\omega_{2}$ to be be determined.
Similar to the FMS scheme, the sum of the energy-momenta of the $\xi\eta$, $\phi_{i}$,
and $\chi_{i}$ CFTs is equivalent to that of the $\beta\gamma$ CFT,
\begin{equation}\label{eq: y11}
T_{\beta\gamma}(z) \cong \sum_{i=1}^{2} [T_{\phi_i}(z) + T_{\chi_i}(z)] +
T_{\xi\eta}(z).
\end{equation}

Now we determine the constants $\omega_{1}$ and $\omega_{2}$. The central charges of
the newly introduced CFTs are
\begin{equation}\label{eq: y12}
\left. \begin{array}{ll} c_{\phi_1} = c_{\phi_2} = 13\,, \,\,\,& c_{\chi_1} =
1-12\omega_1^2\,, \\
c_{\chi_2}=1-12\omega_2^2 \,, & c_{\xi\eta}=-2 \,.
\end{array}
\right. \end{equation} 
Their sum should reproduce the central charge $c_{\beta\gamma}=11$. 
This yields the first constraint on $\omega_{1}$ and $\omega_{2}$,
\begin{equation}\label{eq: y13}
\omega^{2}_{1} + \omega_{2}^{2} = \frac{5}{4}.
\end{equation}
To reproduce the correct conformal dimensions of $\beta(z)$ and $\gamma(z)$,
one needs,
\begin{equation}\label{eq: y16}
\omega_{1} + \omega_{2} = - \sqrt2 \,.
\end{equation}
The solution to Eqs.(\ref{eq: y13}) and (\ref{eq: y16}) is unique,
\begin{eqnarray}\label{eq: y17}
\omega_1 = - \frac{1}{4} \sqrt{2} , \,\,\,\ \omega_2 = - \frac{3}{4} \sqrt{2}.
\end{eqnarray}
This completes our bosonization proposal.

{\it Super-conformal Ghost Current.} For completeness, we re-express the
super-conformal ghost current $J_{\beta\gamma}(z)=:\beta(z)\gamma(z):$ in the new
bosonization scheme
\begin{eqnarray}\label{eq: y18}
J_{\beta\gamma}(z)  \cong
\partial\phi_{1}(z) + \partial\phi_{2}(z) + \frac{1}{\sqrt2}[\partial\chi_{1}(z) +
\partial\chi_{2}(z)] \, .
\end{eqnarray}
This can be easily checked by calculating the OPE of $\beta(z)\gamma(-z)$. In terms of
the original ghost fields $\beta$ and $\gamma$, one has
\begin{equation}\label{eq: y19}
\beta(z)\gamma(-z) = -\frac{1}{2z} + :\beta(0)\gamma(0): + O(z) \, .
\end{equation}
In terms of the bosonized fields, one has alternatively,
\begin{equation}\label{eq: 21}
\left. \begin{array}{ll}  \beta(z)\gamma(-z) \cong  - \frac{1}{2z} & + \{\,
\partial\phi_{1}(0) + \partial\phi_{2}(0)  \\
 & + \frac{1}{\sqrt2}\,[\, \partial\chi_{1}(0) + \partial\chi_{2}(0) \,]\, \} + O(z) \, .
\end{array}
\right.
\end{equation}
The equivalence between Eqs.(\ref{eq: y19}) and (\ref{eq: 21}) ensures the validity of
Eq.(\ref{eq: y18}). The super-conformal ghost current is
actually equivalent to a special algebraic sum of $\phi-\chi$ momentum currents.
For a consistent check, we calculate the following OPEs in terms of the bosonized fields,
\begin{eqnarray}\label{eq: 22}
J_{\beta\gamma}(z)\beta(0) & =   &\frac{1}{z}\beta(0) + O(1), \nonumber \\
J_{\beta\gamma}(z)\gamma(0)& = - &\frac{1}{z}\gamma(0) + O(1).
\end{eqnarray}
This is just what to be expected, the ghost numbers of $\beta$ and $\gamma$ are $1$
and $-1$, respectively.

{\it Vertex Operators for Massless Open String States.} \,\, 
Now we are ready to construct the vertex operators for massless open string states in
intersecting D-brane scenarios. 
For concreteness, we consider open strings connecting two 2-branes
which were described above Eq.(\ref{eq: 1}). 
The D-brane configuration preserves part of the spacetime supersymmetries in open string sectors. 
The vertex operators are of the forms
\begin{eqnarray}\label{eq: 23}
& & V^{(1)}_{B}(z)= e^{[-i(1-\alpha)H_{3}(z)-i\alpha H_{4}(z)]}\Theta_{gh}^{NS}(z) \nonumber\\
& & V^{(2)}_{B}(z)= e^{[i\alpha H_{3}(z)+i(1-\alpha) H_{4}(z)]}\Theta_{gh}^{NS}(z)
\end{eqnarray}
in the NS sector\cite{Douglas} and
\begin{eqnarray}\label{eq: 24}
& V_{F}(z)= & e^{\,[\,is_{0}H_{0}(z) + is_{1}H_{1}(z) + is_{2}H_{2}(z)\,]\,} \nonumber \\
&  & \cdot e^{\,[\,i(\alpha-\frac{1}{2})H_{3}(z) +i(\frac{1}{2}-\alpha) H_{4}(z)\,]\,}
\cdot \Theta_{gh}^{R}(z)
\end{eqnarray}
in the R sector (where $s_{0}, s_{1}, s_{2}=\pm \frac{1}{2}$). 
The superghost factors of the vertex operators obey the OPEs\cite{Pol},
\begin{eqnarray}\label{eq: 25}
&  & \gamma(z)\Theta_{gh}^{NS}(0)= O(z)\,,\, \,\,\,\,\,\,
\beta(z)\Theta_{gh}^{NS}(0)=O(1/z)\,. \\
&  & \gamma(z)\Theta_{gh}^{R}(0)= O(\sqrt z)\,, \,\,\,\,
\beta(z)\Theta_{gh}^{NS}(0)=O(1/{\sqrt z}) \,.
\end{eqnarray}
These imply the general forms
\begin{eqnarray}\label{eq: 27}
& & \Theta_{gh}^{NS}(z)= e^{\,[\,-(1-\alpha)\phi_{1}(z)-\alpha
\phi_{2}(z) + \rho \chi_{1}(z)-\rho \chi_{2}(z)\,]}\,, \nonumber \\
& & \Theta_{gh}^{\,R\,\,\,}(z)= e^{\,[\,-(\frac{1}{2}-\alpha)\phi_{1}(z)-\alpha
\phi_{2}(z) + \kappa \chi_{1}(z)-\kappa \chi_{2}(z)\,]}\,.
\end{eqnarray}
for the vertex operators of the ghost ground states.
For $\rho= 0, \,\, 1/{\sqrt 2}$ and $\kappa= (1 \pm \sqrt {4\alpha+1})/2\sqrt 2$, 
the full vertex operators do have the required conformal dimension of {\it one}.
As expected,
the fermion vertex operators $V_{F}(z)$ in Eq. (\ref{eq: 24}) form a spinor of the Lorentz group $SO(1,5)$. 
GSO projections can be implemented by imposing mutual
locality of the vertex operators\cite{Douglas,Pol}. 
This is automatically satisfied by vertex operators in Eq. (\ref{eq: 23}) in the NS sector. 
In the R sectors, this requires $s_0 + s_1 + s_2 = 1/2$. 
Taking into account the physical state condition $s_0=1/2$, 
we have the following vertex operator for left-handed massless open
string state in the R sector ($s_1=-1/2$),  
\begin{eqnarray}\label{eq: 27-1}
V_{F}(z)&=&\exp{\{ \frac{i}{2} \,[\,H_0(z) -H_1(z) + H_2(z)\,]\, + i(\alpha - \frac{1}{2}) H_3(z) } \nonumber \\
& &{+ i(\frac{1}{2}-\alpha )H_4(z) \}} {\cdot \exp[-(\frac{1}{2}-\alpha )\phi_1(z) -\alpha_2\phi_2(z)} \nonumber \\
& &{+ \kappa \chi_1(z) -\kappa \chi_2(z)]}
\end{eqnarray}

Within the BRST quantization, one can define a nilpotent charge,
\begin{eqnarray}\label{eq: 27-3}
Q & = & \frac{1}{2\pi i} \oint dz \{\, c(z)T^{M}_{B}(z)
+ \gamma(z)T^{M}_{F}(z) \nonumber \\
&& + b(z)c(z)[\,\partial c(z)\,] + \frac{3}{4}\,[\,\partial c(z)\,]\,\beta(z)\gamma(z) \nonumber \\
&& + \frac{1}{4}c(z)\,[\,\partial \beta(z) \,]\,\gamma(z) 
  - \frac{3}{4}c(z)\beta(z)\,[\,\partial \gamma(z)\,]\, \nonumber \\
&& - b(z)\gamma^{2}(z) \,\}\,.
\end{eqnarray}
where $b$, $c$ are the usual conformal ghosts,
$T^{M}_{B}$ is the energy-momentum tensor of matter fields and 
$T^{M}_{F}$ its superconformal partner.
The physical states form a BRST cohomology class,
\begin{equation}\label{eq: 27-2}
Q|\textrm{Phys}>=0\,,
\end{equation}
It is straightforward to verify that
\begin{equation}\label{eq: 27-4}
\,[\, Q, \, \frac{1}{2 \pi i} \oint dz V_{B,F}(z) \,]\,=0.
\end{equation}
So, if $|\Phi>$ is a physical state, $\frac{1}{2\pi i}\oint dz V_{F,B}(z)|\Phi>$
is a physical state as well.
$V_{B,F}(z)$ defined in this letter do indeed satisfy all conditions as required.

It should pointed out that our construction is quite general and can be extended (at
least to cases where there are some unbroken supersymmetries in open string sectors).
For illustration, we present the results for the supersymmetric ``$n=3$'' case in the sense of \cite{Douglas},
where D-branes intersecting with two independent angles $\alpha_2$ and $\alpha_3$. 
The vertex operators for massless open string states are found to be
\begin{eqnarray}\label{eq: 28}
V_{B}(z)&=&\exp{\{ i [ \alpha_2 H_2(z) + \alpha_3 H_3(z) + (1-\alpha_2 -
\alpha_3)H_4(z)] \}} \nonumber \\
& &\cdot \exp[-(1-\alpha_2 - \alpha_3)\phi_1(z) -(\alpha_2 + \alpha_3)\phi_2(z) \nonumber \\
& &+ \rho \chi_1(z) -\rho \chi_2(z)]
\end{eqnarray}
in the NS sector and
\begin{eqnarray}\label{eq: 29}
V_{F}(z)&=&\exp{\{ \pm \frac{i}{2} \,[\,H_0(z) +H_1(z)\,]\,+ i (\alpha_2 -
\frac{1}{2})
H_2(z)  } \nonumber \\
& &{+ i(\alpha_3 - \frac{1}{2}) H_3(z) + i(\frac{1}{2}-\alpha_2 -
\alpha_3)H_4(z) \}} \nonumber \\
& &\cdot \exp[-(\frac{1}{2}-\alpha_2 - \alpha_3)\phi_1(z) -(\alpha_2 + \alpha_3)\phi_2(z) \nonumber \\
& &+ \kappa \chi_1(z) -\kappa \chi_2(z)]
\end{eqnarray}
in the R sector. The condition $h=1$ for these full vertex operators is satisfied if $\rho$
and $\kappa$ are taken as
\begin{eqnarray}\label{eq: 30}
& & \rho=\frac{1}{2\sqrt 2}\,(\,1 \pm
\sqrt{1+8\alpha_2\alpha_3}\,\,)\,, \nonumber\\
& & \kappa=\frac{1}{2\sqrt 2}\,[\,1 \pm \sqrt{4(\alpha_2 + \alpha_3 ) +
8\alpha_2\alpha_3 +1}\,\,]\,.
\end{eqnarray}
These vertex operators commute with the BRST charge.
Furthermore, the fermion vertex operators form a spinor of the Lorentz group $SO(1,\,3)$. 
GSO projections can be carried out by imposing mutual locality of these vertex
operators.

In conclusion, we have constructed a new set of vertex operators for
supersymmetric open string states obeying general boundary conditions. 
The prescription relies on an alternative bosonization of the $\beta\gamma$ CFT, 
in which four independent holomorphic free bosons have been used, 
apart from the conventional anti-commutating $\xi\eta$ ghosts. 
The new bosonization extends to the
fundamental OPEs, the energy-momentum tensor, the super-conformal ghost current, and
thus all correlation functions. 
Although the new scheme looks slightly complicated, it is very powerful and 
provides sufficient rooms for accommodating the requirements to construct reliable
vertex operators for open string states.
The procedure can be readily extended to non-supersymmetric cases\cite{yang}.

\begin{acknowledgments}
HXY would like thank D. Bailin and M. E. Angulo for valuable discussions.
This work is supported in part, by the Pao's Fundation, 
the Startup Foundation of the Zhejiang Education Bureau,
a Fund for Trans-Century Talents, CNSF-90103009 and CNSF-10047005.  
\end{acknowledgments}

\end{document}